\documentclass[twocolumn,showpacs,preprintnumbers,amsmath,amssymb]{revtex4}

\usepackage{mathrsfs}
\usepackage{amsmath}
\usepackage{amssymb}
\usepackage[dvipdfm]{graphicx}
\usepackage{bm}

\usepackage{dcolumn}
\usepackage{bm}

\begin{document}

\preprint{KUNS-2084}

\title{Fate of Kaluza-Klein Black Holes: Evaporation or Excision?}

\author{Keiju Murata}
\email{murata@tap.scphys.kyoto-u.ac.jp}
\author{Jiro Soda}%
 \email{jiro@tap.scphys.kyoto-u.ac.jp}
\author{Sugumi Kanno}
\email{sugumi@hep.physics.mcgill.ca}
\affiliation{
  $^{\ast, \dagger}$Department of Physics,  Kyoto University, Kyoto 606-8501, Japan\\
  $^\ddagger$Department of Physics,  McGill University, Montr\'{e}al, QC H3A 2T8, Canada
}%

\date{\today}


\begin{abstract}
We study evaporation process of black strings which are typical examples of 
Kaluza-Klein black holes.
Taking into account the backreaction of the Hawking radiation, 
we deduce the evolution equation for the radion field. By solving the evolution equation, 
we find that the shape of the internal space is necked by the Hawking radiation and
the amount of the deformation becomes large as the evaporation proceeds. 
 Based on this analysis,
 we speculate that the Kaluza-Klein black holes would be excised from the Kaluza-Klein
  spacetime before the onset of the Gregory-Laflamme instability and therefore before the 
  evaporation. 
\end{abstract}

\pacs{04.70.Dy}
\maketitle

\section{Introduction}

In the conventional 4-dimensional spacetime,
a black hole with the horizon size $r_H$
is considered as a blackbody which emits the Hawking radiation~\cite{Hawking:1974sw}
with the luminosity
\begin{eqnarray}
   L \sim T_{BH}^2
\end{eqnarray}
where $T_{BH} = 1/4\pi r_{H}$ is the temperature of the black hole. 
Because of this radiation, the size of the black hole horizon gradually shrinks
as
\begin{eqnarray}
   \frac{dr_{H}}{dt} \sim - \frac{G}{r_H^2}  \ ,
\end{eqnarray}
where $G$ is the Newton constant.
It is believed that the black hole  evaporates eventually. 

From the point of view of the superstring theory,  it is natural to consider
the higher dimensional spacetime. In the higher dimensional spacetime, 
the event horizon of the black hole would be
the direct product of the usual event horizon and the compact internal
space.
The size of internal space could be much larger than the Planck length
in the large extra dimension scenario~\cite{Arkani-Hamed:1998rs}.
The internal space must be stabilized with some mechanism so that the
theory does not contradict experiments. 
Therefore, naively, we envisage the same picture as the 4-dimensional one for the evaporation
process. 

However, things are not so simple.
Let us consider a black string, which is the simplest example of Kaluza-Klein
black holes, with a circle $S^1$ as the compact internal space.
The evaporation processes of the black strings could be different from
the 4-dimensional one due to the Gregory-Laflamme instability~\cite{Gregory:1993vy}
( see also \cite{Harmark:2007md} and references therein). 
In fact, it is known that the black string is unstable when the horizon radius is
smaller than the length scale of compactification. This
instability changes the spacetime structure.

\begin{figure}[h]
\begin{center}
\includegraphics[height=5cm,clip]{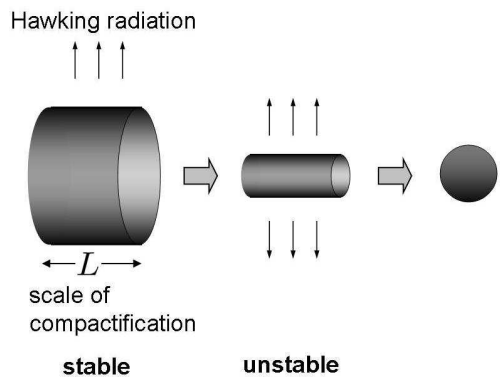}
\caption{\label{fig:naiveKKBHevap} A picture of the evaporation process of the
 black string. Due to the Hawking radiation, the black string becomes thin.
 Because of the Gregory-Laflamme instability, the black string is transformed
  to  the spherically  symmetric black hole. Subsequently, the black hole evaporates. }
\end{center}
\end{figure}

Soon after the onset of the Gregory-Laflamme instability, 
it is believed that the higher dimensional spherically symmetric black hole, the so-called 
Myers-Perry black hole~\cite{Myers:1986un}, is formed.
Once this transition is assumed, the subsequent process is the
5-dimensional one.
The evaporation
process taking into account this instability is
depicted in Fig. \ref{fig:naiveKKBHevap}.
However, the detail of the spacetime dynamics after
the Gregory-Laflamme instability is not well known. Hence, we will only consider the
evaporation process before the Gregory-Laflamme instability in this
paper.

Moreover, the dynamics of the internal space, the so-called radion,
 can not be neglected all the time. 
It is reasonable to assume that the radion is constant at low energy. 
However, for sufficiently small black holes, 
this would not be true. Indeed, when the mass scale of the stabilized radion
 becomes less than $1/r_H$, the radion is effectively free. 
Hence, in contrast to the conventional assumption that the radion is always fixed
during the evaporation process, we take an attitude that the radion is 
crucial in understanding of the evaporation
process of the black string. In fact,
the backreaction of the Hawking radiation could change the radion. 
It is therefore natural to expect the radion plays an important role
in the evaporation process of the black strings.
Previously, we have investigated the interplay between the radion and 
the Gregory-Laflamme instability~\cite{Kanno:2003au,Tamaki:2003bq,Kanno:2005mv}.
 To the best of our knowledge,
however, the interplay between the Hawking radiation and  
the radion dynamics has not been considered at all
(except for our preliminary report \cite{Murata:2007vp}).

In this paper, we study the role of the radion dynamics
in the evaporation process of the black string and show that  
  the internal space is deformed inhomogeneously by the Hawking radiation. 
  Based on this result, we can speculate the fate of the Kaluza-Klein
  black holes as follows. The black hole shrinks due to Hawking radiation.
  At the same time, the internal space is necked.
  This process continues quasi-stationary.  And, the spacetime is eventually 
  pinched at the radius close to the black hole horizon.
  Consequently, assuming the resolution of the singularity, 
 the black string would be excised from the spacetime. 
We stress that the black string excision from the spacetime
could occur before the evaporation through the conventional process.

The organization of this paper is as follows. In section II, we introduce the Kaluza-Klein
black hole solutions. In order to make the problem tractable, we do dimensional reduction
to 2-dimensional problem. In section III, we derive the evolution equation for the radion
field with the backreaction of the Hawking radiation. In section IV, we analyze
the radion dynamics and speculate the fate of the Kaluza-Klein black holes.
 The final section is devoted to the conclusion.

\section{dimensional reduction of Kaluza-Klein black holes}

In this section, first of all, 
we introduce the black string solutions as the simplest example 
of the Kaluza-Klein black holes. 
Then, we show that the black string system can be approximated by
the 2-dimensional effective action
 where we can easily incorporate the backreaction
of the Hawking radiation. 

 Let us consider the $D$-dimensional Einstein gravity 
 with a massless scalar field $f$,
\begin{multline}
 S = M_D^{D-2} \int d^Dx\sqrt{-g_{(D)}}\, R^{(D)}\\+ \int
  d^Dx\sqrt{-g_{(D)}}\, \left[-\frac{1}{2}(\nabla f)^2 \right]\ ,
\label{eq:E-H}
\end{multline}
where $g^{(D)}_{MN}$ ($M,N=0,1,\dots,D-1$) and $R^{(D)}$ are $D$-dimensional metric and Ricci
scalar, respectively. Here, $M_D$ denotes the $D$-dimensional Planck mass.
The equations of motion of the action (\ref{eq:E-H}) are given by
\begin{eqnarray}
 G_{MN} &=& \frac{1}{M_D^{D-2}} \left[\nabla_M f \nabla_N f 
- \frac{1}{2}g_{MN}(\nabla f)^2 \right]\ ,\\
\Box f &=& 0 \ .
\end{eqnarray}
In the vacuum case $f=0$, 
it is easy to find a black string solution, which is an example of
Kaluza-Klein black holes. The result is given by
\begin{eqnarray}
 ds^2 &=& -Vdt^2
 + V^{-1}dr^2
  + r^2 d\Omega_{n-2}^2 + dy^2\ ,   \nonumber\\
V&\equiv & 1-\left(\frac{r_H}{r}\right)^{n-3}   , \quad  n \geq 4 \ , 
\label{eq:bs}
\end{eqnarray}
where $d\Omega_{n-2}^2$ is the line element of 
the $(n-2)$-dimensional sphere $S^{n-2}$.
Here, we defined $n=D-1$, that is, $n$ is the dimension of the 
external space. The direction of $y$ is compactified with the length $L$. 
We would like to consider the evaporation process of this black string. 
Unfortunately, it is difficult to perform  semi-classical
analysis without any approximation.
Therefore, we reduce the degrees of freedom of the system
by dimensional reduction.  After that, we quantize the system.

First of all, we show that the system can be reduced to the 3-dimensional
one where we can show the existence of the Gregory-Laflamme instability.
The black string spacetime (\ref{eq:bs}) has $SO(n-1)$ symmetry. We
assume the evaporation proceeds keeping this symmetry. In other words,
we assume the functional form of the scalar field as $f=f(x^\mu)$ and 
the metric as
\begin{equation}
 ds^2 = g^{(3)}_{\mu\nu}(x^\mu)dx^\mu dx^\nu +
  M_D^{-2}e^{-\frac{4}{n-2}\phi(x^\mu)}d\Omega_{n-2}^2\ ,
\end{equation}
where $\mu,\nu=0,1,2$. 
Then, in the action (\ref{eq:E-H}), we can carry
out the integration of angular variables and the result becomes
\begin{multline}
 S = \omega_{n-2}M_D \int d^3x \sqrt{-g_{(3)}}\, e^{-2\phi}\\
\times \left[ R^{(3)} +
 4\frac{n-3}{n-2}(\nabla\phi)^2 +
 M_D^2(n-2)(n-3)e^{\frac{4}{n-2}\phi} \right]  \\
+ \int d^3x \sqrt{-g_{(3)}}\,e^{-2\phi} \left[ -\frac{1}{2}(\nabla f)^2 \right]\ ,
\label{eq:3Daction}
\end{multline}
where $R^{(3)}$ is the 3-dimensional Ricci scalar and $\omega_{n-2}$ is the
area of $(n-2)$-dimensional unit sphere,
\begin{equation}
 \omega_{n-2} = \frac{2\pi^{(n-1)/2}}{\Gamma((n-1)/2)}\ .
\end{equation}
Thus, we have obtained the 3-dimensional effective action. 
Note that the formal limit $n \rightarrow \infty$ gives the model similar to
that proposed by Callan et al~\cite{Callan:1992rs} as is shown 
in \cite{Soda:1993xc}.
We have numerically  analyzed the classical stability using this action and
confirmed the existence of
 the Gregory-Laflamme instability~\cite{Gregory:1993vy,Harmark:2007md}. 

Let us further reduce the system 
into the 2-dimensional one where we can easily incorporate the backreaction
of the Hawking radiation. 
 We assume that the horizon radius $r_H$ is much larger than the length scale of
compactification $L$. In that case,
 the Gregory-Laflamme instability is not relevant. 
and, hence, we can ignore the Kaluza-Klein modes.
Then we can take the functional form of the scalar field as $f=f(x^a)$
 and  the metric as
\begin{equation}
 g^{(3)}_{\mu\nu}dx^\mu dx^\nu = g_{ab}(x^a)dx^a dx^b +
  e^{-2\chi(x^a)}dy^2\ ,
\end{equation}
where $a,b=0,1$. The radion is denoted by $\chi$, which describe
the dynamics of the internal space. Then, in the
3-dimensional action (\ref{eq:3Daction}), we can carry out the 
integration of $y$. The resultant action is
\begin{multline}
 S = \omega_{n-2}M_D L\int d^2x \sqrt{-g}\, e^{-2\phi -\chi}\\
  \times \left[ R +
 4\frac{n-3}{n-2}(\nabla\phi)^2 + 4\nabla \phi \cdot \nabla \chi  \right. \\
\left.  + (n-2)(n-3)M_D^2e^{\frac{4}{n-2}\phi} \right] \\
+ \int d^2x \sqrt{-g}\,e^{-2\phi-\chi} \left[-\frac{1}{2}(\nabla f)^2 \right]\ ,
\label{eq:2Daction}
\end{multline}
where $R$ is the 2-dimensional Ricci scalar.
Here, we have rescaled the field $f$ to absorb the volume factor.
 The solutions of the equations of motion derived from
 the action (\ref{eq:2Daction}) can be read off from
the black string solution (\ref{eq:bs}) as
\begin{gather}
  g_{ab}dx^a dx^b = -Vdt^2+ V^{-1}dr^2\ ,\label{eq:bg1}\\
 M_D^{-1}e^{-\frac{2}{n-2}\phi}=r\ ,\label{eq:bg2}\\
 \chi = 0\ ,\label{eq:bg3}\\
 f=0\ .\label{eq:bg4}
\end{gather}
We shall study the evaporation process of the black string
with the above 2-dimensional action (\ref{eq:2Daction}).
Of course, the approximation we performed may not
give the quantitatively correct answer. However, we believe the qualitative result 
will not be altered.

\section{Backreaction of Hawking Radiation}

In this section, we derive the master equation for the radion field with
 the backreaction of the Hawking radiation.

The Hawking radiation is the quantum effect of the scalar field in
Eq. (\ref{eq:2Daction}), 
\begin{equation}
 S_\text{m}[g,\phi,\chi,f] = \int d^2x
  \sqrt{-g}\,e^{-2\phi-\chi} \left[ -\frac{1}{2}(\nabla f)^2 \right] \ .
\label{eq:matt}
\end{equation}
The other fields are treated as classical ones. This semiclassical
approximation can be implemented by considering the effective action
\begin{equation}
 W[g,\phi,\chi] = -i \ln\left[\int \mathscr{D}f\,
			 e^{iS_\text{m}[g,\phi,\chi,f]}\right]\ .
\label{new-matter}                   
\end{equation}
What we need to do is to replace the action (\ref{eq:matt})
by the above effective action (\ref{new-matter}) 
and  derive equations of motion. 
Let us  define 
\begin{gather}
 T_{ab} \equiv \frac{-2}{\sqrt{-g}}\frac{\delta W}{\delta
 g^{ab}}\ ,\\
 X \equiv \frac{1}{\sqrt{-g}}\frac{\delta W}{\delta \phi}\
 ,\label{eq:Xdef}\\
 Y \equiv \frac{1}{\sqrt{-g}}\frac{\delta W}{\delta \chi}\ .
\label{eq:Ydef}
\end{gather}
Fortunately, the explicit functional forms of $X$ and $Y$ are not necessary,
because they disappear in the final equation. 
The variation with respect to the metric gives
\begin{multline}
  2\nabla_a\nabla_b \phi
+ \nabla_a\nabla_b \chi \\
- \frac{4}{n-2}\nabla_a\phi \nabla_b\phi 
- \nabla_a\chi \nabla_b \chi
+ g_{ab} \{-2\nabla^2 \phi \\ 
- \nabla^2\chi +  2\frac{n-1}{n-2}(\nabla\phi)^2
+ 2\nabla\phi\cdot\nabla\chi + (\nabla\chi)^2  \\
- \frac{1}{2}(n-2)(n-3)M_D^2e^{\frac{4}{n-2}\phi} \}
             =\kappa  e^{2\phi +\chi} T_{ab}  \ ,
\label{eq:delg}
\end{multline}
where $\kappa^{-1} = 2 \omega_{n-2}M_D L $.  
Notice that the Einstein tensor in 2-dimensions vanish. 
From the variation with respect to the dilaton $\phi$, we obtain 
\begin{multline}
 R + 4\frac{n-3}{n-2}\nabla^2\phi +
 2\nabla^2 \chi\\ - 4\frac{n-3}{n-2}(\nabla\phi)^2 - 4\frac{n-3}{n-2}\nabla\phi\cdot\nabla\chi
 -2(\nabla\chi)^2 \\+ (n-3)(n-4)M_D^2e^{\frac{4}{n-2}\phi} = \kappa e^{2\phi +\chi} X \ ,
\label{eq:delphi}
\end{multline}
and, finally, from the variation with respect to the radion $\chi$, we have 
\begin{multline}
 R + 4\nabla^2 \phi -4\frac{n-1}{n-2}(\nabla\phi)^2\\
 + (n-2)(n-3)M_D^2e^{\frac{4}{n-2}\phi}
 = 2 \kappa e^{2\phi +\chi} Y \ .
\label{eq:delchi}
\end{multline}
The above five equations determines the five variables $g_{ab} , \phi$ and $\chi$.

Now, we derive the evolution equation for the radion. 
The trace of Eq. (\ref{eq:delg}) becomes
\begin{multline}
 -2\nabla^2 \phi - \nabla^2 \chi + 4(\nabla\phi)^2 + 4\nabla\phi\cdot
  \nabla\chi + (\nabla\chi)^2\\ - (n-2)(n-3)M_D^2e^{\frac{4}{n-2}\phi} =
 \kappa e^{2\phi+\chi}T^a_a\ .
\label{eq:1}
\end{multline}
From Eq. (\ref{eq:delphi}) and (\ref{eq:delchi}), we can eliminate the curvature
as
\begin{multline}
 -2\nabla^2 \phi + (n-2)\nabla^2 \chi +
  4(\nabla\phi)^2 - 2(n-3)\nabla\phi \cdot
  \nabla\chi\\ -(n-2)(\nabla\chi)^2 -
 (n-2)(n-3)M_D^2e^{\frac{4}{n-2}\phi}\\
 =\frac{(n-2) \kappa}{2} e^{2\phi+\chi}(X-2Y)\ .
\label{eq:2}
\end{multline}
From Eq. (\ref{eq:1}) and (\ref{eq:2}), we obtain
\begin{multline}
 (\nabla\chi)^2 - \nabla^2\chi + 2\nabla\phi\cdot\nabla\chi \\=
  \frac{(n-2)\kappa }{2(n-1)} e^{2\phi+\chi}(\frac{2}{n-2}T^a_a-X+2Y)\ .
  \label{nonlinear}
\end{multline}
The fields $\phi$ and $\chi$ appear in the classical action (\ref{eq:matt})
only in the combination $2\phi+\chi$, $S_\text{m}=S_\text{m}[g,f,2\phi+\chi]$.
 Hence, they have to come into the effective action with the same
 combination, $W = W[g,2\phi+\chi]$. Then,
from Eq. (\ref{eq:Xdef}) and (\ref{eq:Ydef}), we get the relation $X = 2Y$. 
Thus, from Eq.(\ref{nonlinear}), we obtain the equation
\begin{multline}
 (\nabla\chi)^2 - \Box\chi + 2\nabla\phi\cdot\nabla\chi =
  \frac{\kappa}{(n-1)}e^{2\phi+\chi}T^a_a \ .
\label{eq:chiNL}
\end{multline}

As Eq.(\ref{eq:chiNL}) is the nonlinear equation, it is difficult to
solve it exactly.  
Now, we consider the case $r_H \gg M_D^{-1}$. 
That implies the Hawking temperature $T_{BH}$ is low and
the Hawking flux is weak.
Then, we can treat the back reaction of the Hawking radiation as the
perturbation and regard $T_{ab}$ as the source of perturbation. We take
Eqs. (\ref{eq:bg1}) $\sim$ (\ref{eq:bg3}) as the background solution.
 Then, Eq. (\ref{eq:chiNL}) becomes
\begin{equation}
 -\Box \delta\chi + 2\nabla\phi \cdot \nabla\delta\chi 
= \frac{\kappa}{(n-1)}e^{2\phi}T^a_a\ .
\label{eq:master}
\end{equation}
Here, it should be noted that $\phi$ takes the background value. 
The above Eq.(\ref{eq:master}) is a master equation 
for the radion perturbation $\delta\chi$. 
Note that the perturbed radion is gauge invariant since the background vanishes, $\chi=0$. 
Therefore, the gauge mode cannot
appear in the master equation (\ref{eq:master}).
It is worth to note that, the master equation
(\ref{eq:master}) is a wave equation with a source term. 
Since the classical action of the
matter field (\ref{eq:matt}) is Weyl invariant, the trace part of the
energy-momentum tensor $T^a_a$ should be zero classically. 
However, Weyl symmetry has the anomaly 
in the quantum theory. The trace anomaly is well known and given by
\begin{equation}
 T^a_a = \frac{R}{24\pi}    \ .
\label{eq:traceanom}
\end{equation}
It should be stressed that this trace anomaly is intimately related to 
the Hawking radiation~\cite{Christensen:1977jc}.
Thus, it turned out that the Hawking radiation induces the radion dynamics.

\section{Fate of Kaluza-Klein Black Holes}

In this section, we analyze the master equation (\ref{eq:master}) 
and show how the radion is deformed.
 Based on this linear analysis, we speculate the fate of the black strings.

In the Schwarzschild coordinates $(t,r)$, the master equation
(\ref{eq:master}) becomes 
\begin{equation}
 \delta\chi_{,tt}-V^2\delta\chi_{,rr} +
  (2V\phi_{,r}-V_{,r})V\delta\chi_{,r} = F(r)\ ,
\label{eq:master(tr)}
\end{equation}
where $F(r)$ is the  external force,
\begin{equation}
 F(r)\equiv \frac{\kappa}{(n-1)}Ve^{2\phi}T^a_a
        = - \frac{\kappa}{24(n-1)\pi}e^{2\phi} V V_{,rr}    \ .
\label{eq:Fdef}
\end{equation}
It is convenient to define the tortoise coordinate,
\begin{equation}
 r_\ast = \int \frac{dr}{V}\ ,
\end{equation}
and rewrite  the master equation (\ref{eq:master}) as
\begin{equation}
 \delta\chi_{,tt} - \delta\chi_{,\ast\ast} +
  2\phi_{,\ast}\delta\chi_{,\ast} = F(r_\ast)\ ,
\label{eq:master_r_ast}
\end{equation}
where $_{,\ast}\equiv \partial/\partial r_\ast$ and 
$F(r_\ast)\equiv F(r(r_\ast))$. 

It is useful to see the explicit functional form of the source term. 
Notice that the asymptotic forms of the $r_\ast(r)$ is
\begin{equation}
r_\ast \sim 
\begin{cases}
 r & (r \rightarrow \infty)\\
 \frac{1}{n-3}r_H \ln\left(\frac{r-r_H}{r_H}\right) &
 (r \sim r_H)\ ,
\end{cases}
\end{equation}
and its inverse function $r(r_\ast)$ is given by
\begin{equation}
\begin{cases}
 r \sim r_\ast & (r_\ast \rightarrow \infty)\\
 \frac{r-r_H}{r_H} \sim e^{(n-3)r_\ast/r_H} &
 (r_\ast \rightarrow -\infty)\ .
\label{eq:r(r_ast)}
\end{cases}
\end{equation}
Hence, from Eq. (\ref{eq:Fdef}) and
(\ref{eq:r(r_ast)}), the asymptotic form of the $F(r_\ast)$ can be deduced as
\begin{multline}
 F(r_\ast) \sim \frac{(n-2)(n-3)\kappa }{24 \pi (n-1) M_D^{n-2}} 
\frac{1}{r_\ast^n}\left(\frac{r_H}{r_\ast}\right)^{n-3}\\
(r_\ast \rightarrow \infty)\ ,
\end{multline}
and 
\begin{multline}
 F(r_\ast) \sim \frac{(n-2)(n-3)^2 \kappa}{24\pi
  (n-1)M_D^{n-2}\,r_H^n}
e^{(n-3)r_\ast/r_H}\\
(r_\ast \rightarrow -\infty)\ .
\end{multline}
\begin{figure}
\begin{center}
\includegraphics[height=4cm,clip]{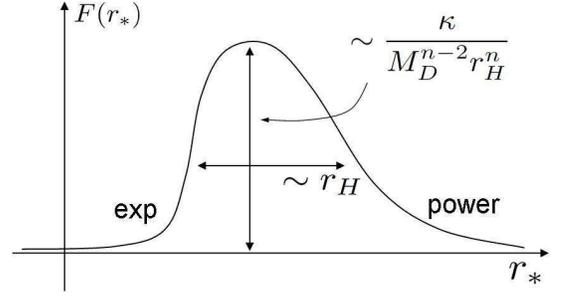}
\caption{\label{fig:F(r_ast)}The graph of the
 external force $F(r_\ast)$. This function has the height 
$\sim \kappa /(M_D^{n-2} \,r_H^n)$ and width $\sim r_H$.}
\end{center}
\end{figure}
From the above expression, 
we can depict the figure of the function $F(r_\ast)$
as in Fig.\ref{fig:F(r_ast)}. We can see that this function has the
 height $\sim \kappa /(M_D^{n-2}\,r_H^n)$ and width $\sim r_H$.
The peak is located at the point close to the horizon 
in the original Schwarzschild coordinates.

\begin{figure}[h]
\begin{center}
\includegraphics[height=5cm,clip]{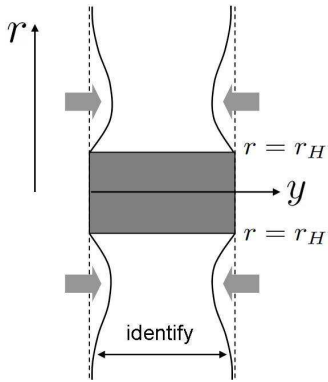}
\caption{\label{fig:radion_dynamics_onset}The radion has inhomogeneous
profile due to the Hawking radiation.}
\end{center}
\end{figure}

Now, we shall analyze the master equation.
The specific choice of initial conditions is not important for our result.
So, let us choose the simplest initial conditions 
\begin{eqnarray}
\delta\chi(t=0,r_\ast )=\delta\chi_{,t}(t=0,r_\ast )=0 .
\end{eqnarray}
Then, from Eq.(\ref{eq:master_r_ast}), we can deduce the radion dynamics
for small $t$, 
\begin{equation}
\begin{split}
 \delta\chi = \frac{1}{2} F(r_\ast ) \ t^2 \ .
\end{split}
\label{eq:radion_dynamics}
\end{equation}
 We gave a schematic picture of the radion dynamics in the 
 Fig.\ref{fig:radion_dynamics_onset}.
However, this is just the initial motion. To see what occurs subsequently,
let us define a new variable $u = e^{-\phi} \delta \chi$.
In terms of the new variable, we can write the master equation 
(\ref{eq:master_r_ast}) as
\begin{equation}
 u_{,tt} - u_{,\ast\ast} + m_{eff}^2  u
                    =  e^{-\phi} F(r_\ast) \ .
                    \label{wave}
\end{equation}
 The effective mass term in Eq.(\ref{wave}) can be calculated as 
\begin{eqnarray}
  m_{eff}^2 &\equiv &  e^{\phi} \left( e^{-\phi} \right)_{,\ast \ast} \nonumber\\
  &=& \left( \frac{n}{2} -1 \right) \frac{V}{r^2}
  \left[ \left( n -3 \right) - \left( \frac{n}{2} -1 \right) V \right]
\end{eqnarray}
The point is that the effective mass $m_{eff}^2$ is positive everywhere.
The order of magnitude can be estimated as  $m_{eff}^2 \sim 1/r_H^2$.
From Eq.(\ref{wave}), we see that the mass term prevent the motion of the 
radion. As a consequence, the radion approaches the static equilibrium solution.
The amount of the deformation can be evaluated as
\begin{eqnarray}
    \delta \chi &=& \frac{ F(r_\ast)}{m_{eff}^2}  \nonumber\\
    &=& \frac{(n-3)\kappa \left(1-V \right) }{6\pi (n-1) (M_D r)^{n-2} 
    \left[2(n-3) - (n-2) V \right]}  \ .  \quad
    \label{deformation}
\end{eqnarray}
The profile of the radion is inhomogeneous as can be seen from
the formula (\ref{deformation}). 
More accurately, we have solved Eq.(\ref{eq:master(tr)}) numerically
and found that there is a maximum very close to the horizon. 
From Eq.(\ref{deformation}), the maximal value of $\delta \chi$ can be estimated as
 $ \delta \chi \sim \left(M_D r_H \right)^{1-n} r_H /L $.
Thus, we found that the backreaction of the Hawking radiation, which makes the 
black hole shrink, also deforms the radion inhomogeneously.

We now speculate the fate of the black string. 
The internal space is pushed  by the Hawking radiation and tend to
collapse. However, the effective mass term sustain the internal space. 
As a result, the internal
space is slightly necked and stabilized there.
In this quasi-stationary phase, the black hole is shrinking
due to the loss of the mass in the Hawking radiation.
As the horizon radius $r_H$ is decreasing, the size of 
the neck is also shrinking adiabatically as $1/r_H^{n-2}$.
Therefore, at the end of the day, the internal space would be
pinched at some radius close to the horizon.
The topology of the resultant singular surface is a $(n-2)$-dimensional sphere. 
Assuming the resolution of the singularity, we can conclude that
 the black string is eventually excised from Kaluza-Klein spacetime.
Therefore, we get the picture sketched in Fig.4. 
Of course, there is a possibility the conventional evaporation 
ends before the excision. 
Even in that case, the excision will occur
and the evaporating region is hidden behind 
the pinched point.
Therefore, this excision process may add a complication to 
the information loss problem in black hole physics~\cite{Preskill:1992tc}.

\begin{figure}[h]
\begin{center}
\includegraphics[height=4cm,clip]{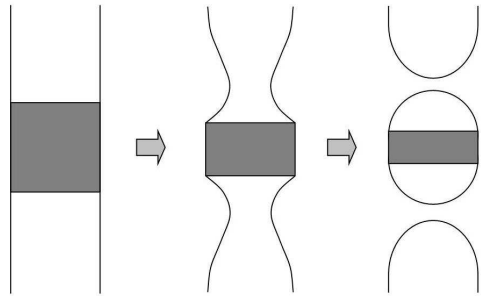}
\caption{\label{fig:radion_dynamics_nonlinear} The speculative
 non-linear radion dynamics is shown. 
 The black string shrinks due to the Hawking radiation. At the same time,
 the radion is deformed by the backreaction of the Hawking radiation.
 The black string would be excised from the spacetime
 at the end of the day. }
\end{center}
\end{figure}

There are several caveats.
Firstly, there may be a Gregory-Laflamme instability before excision
depending on $L$. We have assumed $L$ is sufficiently small.
Secondly, the radion is stabilized in the realistic models, although we
analyzed as if the radion has not been stabilized at all. However, when
Hawking temperature becomes much larger than the scale of
the radion stabilization,  the radion is effectively free.
We are looking at this stage. 
Thirdly, we used the dimensional reduction method to simplify the
problem. However, in quantization, the connection of 2-dimensional
models with a real world becomes unclear due to the existence
of the dimensional reduction anomaly~\cite{Frolov:1999an}. 
For complete analysis, 
we should calculate all of the anomalous terms and 
add them to the 2-dimensional effective action (\ref{new-matter}).
We leave it for future work.
Finally, we may have to worry about the fate of the bubble like 
spacetime after the excision of the black string.
Although we do not have a conclusive answer to this important question,
it would be reasonable to imagine such a tiny hole disappears
 and nothing remains there.

\section{Conclusion}

 We have studied the evaporation process of the black strings
 as the simplest example of Kaluza-Klein black holes
 with focusing on the role of the internal space. 
 We have obtained the master equation for the radion field
 with the backreaction of the Hawking radiation. 
 It turned out that the internal space is deformed inhomogeneously
  (Fig.\ref{fig:radion_dynamics_onset}). Based on this result, we speculated
that the black string would be excised from the spacetime
 due to the non-trivial dynamics of the radion 
 (Fig.\ref{fig:radion_dynamics_nonlinear}). 
 This disappearance of the black string 
  is different from the evaporation process naively considered
so far.  To further confirm the  excision of black strings, 
 we need to analyze the non-linear radion dynamics numerically.

We have considered the Kaluza-Klein black holes in the spacetime 
with the $S^1$ compactification. It is reasonable to believe
 that black holes in the spacetime with the more realistic compactification
  also give the same behavior as that obtained in the present analysis. 
So, it is intriguing to study general Kaluza-Klein black holes in the
string theory.
There may be a relation between our result and recently proposed 
mechanisms
in the context of the closed tachyon condensation~\cite{Horowitz:2005vp}.

We should note that there may exist other different types of Kaluza-Klein black holes. 
 In the context of the braneworld, the localized 
 black hole on the brane would be possible (see the recent review
 \cite{Kanti:2004nr} and references therein). 
 In this context, a very
 interesting excision mechanism of the black hole is discussed in 
\cite{Frolov:2002gf,Frolov:2002as,Flachi:2005hi}. 
 Another possible type of Kaluza-Klein black holes is 
 the squashed Kaluza-Klein black hole~\cite{Ishihara:2005dp}.
 The squashed Kaluza-Klein black hole looks like 5-dimensional black hole in the vicinity
 of the horizon, however, the spacetime far from the black hole
 is locally that of the black string. 
 It is also of interest to consider the evaporation
process of squashed Kaluza-Klein black holes~\cite{Ishihara:2007ni}.

The implication of our result in cosmology also deserve investigations.
In particular,  cosmological consequences of 
the evaporation  of primordial black holes
should be reconsidered in the light of our results.

\vskip 0.2cm
J.S. is supported by  
the Japan-U.K. Research Cooperative Program, the Japan-France Research
Cooperative Program,  Grant-in-Aid for  Scientific
Research Fund of the Ministry of Education, Science and Culture of Japan 
 No.18540262 and No.17340075.  
S.K. is supported by JSPS Grant-in-Aid 
for Research Abroad.

\end{document}